\documentclass[conference,comsoc]{IEEEtran}
\usepackage{xcolor}

\ifCLASSINFOpdf
  \usepackage[pdftex]{graphicx}
  \DeclareGraphicsExtensions{.png}
\else
  \usepackage[dvips]{graphicx}
\fi

\ifCLASSOPTIONcompsoc
  \usepackage[caption=false,font=normalsize,labelfont=sf,textfont=sf]{subfig}
\else
  \usepackage[caption=false,font=footnotesize]{subfig}
\fi

\usepackage{url}

\begin{document}

\title{High-Precision APT Malware Attribution with Out-of-Scope Resilience}

\author{
\IEEEauthorblockN{
Peter Williams\IEEEauthorrefmark{1},
Adam Sobey\IEEEauthorrefmark{1}\IEEEauthorrefmark{2},
Erisa Karafili\IEEEauthorrefmark{1}
}
\IEEEauthorblockA{\IEEEauthorrefmark{1}
University of Southampton, Southampton, UK\\
Email: \{P.G.Williams, ajs502, E.Karafili\}@soton.ac.uk}
\IEEEauthorblockA{\IEEEauthorrefmark{2}
The Alan Turing Institute, London, UK}
}

\maketitle

\begin{abstract}
Early attribution of Advanced Persistent Threat (APT) activity can help defenders prioritise investigation, select countermeasures, and reduce the impact of an intrusion. Malware provides useful attribution evidence, but automated APT malware attribution remains difficult in practice. Existing approaches are typically trained and evaluated as closed-set classifiers over a limited number of known APT groups. In operational environments, however, classifiers are likely to encounter samples from groups not represented during training. Closed-set classifiers are then forced to assign such samples to known groups, producing unsupported and potentially misleading attributions.

We present a high-precision APT malware attribution method based on ranked binary classifiers with explicit abstention. Rather than training a single multi-class classifier, our approach trains and tunes two binary classifiers per APT group, ranks the classifiers by validation performance, and applies them sequentially. A sample is attributed only when a classifier provides sufficient evidence; otherwise, it abstains.
We evaluate the method on the APT Malware dataset and on a larger combined dataset designed to stress-test out-of-scope behaviour. On the APT Malware dataset, the method achieves higher precision than previously published results on the same dataset. In the most challenging setting, where 87\% of test samples came from 60 APT groups excluded from training, the method abstained on 94\% of out-of-scope samples while maintaining 92\% precision and 95\% selective accuracy on the samples it classified.
\end{abstract}


\section{Introduction}
Advanced Persistent Threats (APTs) are very difficult to detect and defend against. Unlike attacks such as Distributed Denial of Service (DDoS), which aim to cause immediate disruption, APTs typically prioritise stealth, persistence and operational secrecy. As a result, identifying the actor behind an intrusion is difficult, particularly when evidence is incomplete, deliberately obfuscated, or only available after the attack has progressed.

The exact number of APTs and their impact is hard to quantify as many are not disclosed publicly, especially those on sensitive targets. Nevertheless, recent industry reporting indicates that APT activity is increasing.
Kaspersky estimates that the number of APTs increased by 74\% between 2023 and 2024 and affected 25\% of companies~\cite{Advanced2024}.
Additional market data suggests that 80\% of attacks remain undetected for at least six months, by which time 48\% of affected organisations have already suffered significant damage. 
It also indicates that 57\% of organisations have insufficient cyber resources, with 65\% believing their tools are insufficient \cite{Advanced2025}. These pressures motivate automated approaches that can assist defenders by reducing the time and expertise required to identify likely adversaries.

Automated attribution is typically treated as a closed-set classification or clustering problem, where a classifier is trained on samples, previously associated with known adversaries by expert analysts, and is then used to classify unseen samples into one of those known groups.
This approach is problematic in operational environments, where defenders may encounter malware from \emph{out-of-scope (OOS)} groups, i.e., APT groups that were not included in the training data. Existing closed-set approaches are typically not designed or evaluated for this setting, limiting their practical applicability. 

In this paper, we propose a novel high-precision machine learning approach for APT malware  attribution based on ranked binary classifiers. Each binary classifier is tuned to recognise malware from a specific APT group. Unlike closed-set multi-class approaches, our method can abstain when no classifier provides sufficient evidence for attribution. This allows the classifier to produce high-precision classifications while identifying samples that require expert human investigation, including samples from groups that were not represented during training.

We first built a multi-class model that outperformed previously published models evaluated on the same dataset~\cite{Xu2021AnLightGBM,Kida2023Nation-StateHashing}. This model was then used as the baseline for subsequent experiments. Since samples from different APT groups may require different model types and tuning strategies, we trained a pair of binary classifiers for each APT group, with the classifier type and tuning selected empirically. Rather than combining the classifiers using a One-vs-Rest approach based on the highest predicted probability, we ranked the classifiers according to our confidence in them and applied them sequentially. Compared with the baseline multi-class model, our ranked binary classifiers reduced incorrect classifications by 52\%, while reducing correct classifications by only 3\%.

We evaluated the resilience of our approach to samples from out-of-scope APT groups using two strategies. For the smaller dataset, we used a leave-one-group-out procedure, where each APT group was treated in turn as out-of-scope by removing its samples from the training set and then evaluating on the full test set. 
For the larger dataset, we trained only on the groups with the most samples and tested against the complete dataset.  
In the most challenging setting, where 87\% of the test samples came from 60 out-of-scope groups, our method abstained on 94\% of out-of-scope samples, demonstrating resilience even when the majority of test samples originate from groups unseen during training.

The main contribution of this paper is an automated APT malware attribution method that combines high-precision classification with explicit abstention. The method achieves higher precision than previously published results on the APT Malware dataset~\cite{Cyber-research/APTMalware:Samples}. When evaluated on a larger dataset in which 87\% of samples originated from groups not represented during training, it abstained on 94\% of out-of-scope samples while maintaining 95\% selective accuracy and 92\% precision on samples from in-scope groups.

This paper is organised as follows. Section \ref{sec:methodology}, details the proposed methodology. Section \ref{sec:metrics} explains the metrics used throughout this paper. Section \ref{sec:results} introduces the exact experiments that were carried out and their results. Section~\ref{sec:discussion}, discussed our findings and observations. Section~\ref{sec:related work}, summarises the related work in the field. While Section~\ref{sec:conclusion and future work}, draws together our conclusions and suggestions for future work.

\section{Methodology}\label{sec:methodology}
In this section, we describe our novel methodology for malware attribution to APT groups.

Our methodology has five main phases:
\begin{enumerate}
    \item \textbf{Dataset Preparation}: We selected and acquired published datasets containing APT malware samples~\cite{Cyber-research/APTMalware:Samples,SecPriv/adapt:Samples}. Features were extracted from the malware samples and converted into a numerical representation.
    \item \textbf{Multi-class and Binary Classifiers}: To provide an in-scope baseline, we trained a set of multi-class machine learning models over the training set.
    We then trained binary classifiers for each APT group which could independently determine whether a previously unseen sample belonged to that specific group.
    \item \textbf{Ranked Binary Classifiers}: The binary classifiers were ranked and combined so that together they could classify whether an unseen sample belonged to any one of the known groups and, if so, which group. Samples that were not accepted by any binary classifier remained unclassified through abstention.
    \item \textbf{Out-of-scope Group Testing}: To provide an out-of-scope baseline, we first tested the baseline multi-class model's resilience to the presence of samples from out-of-scope groups. In each experiment, one group was designated as the out-of-scope group: all of its samples were removed from the training set but retained in the final testing set. This was repeated for all groups and the results were averaged. We then repeated the same experiments using our ranked binary classifiers. 
    \item \textbf{Large Scale Out-of-scope Group Testing:} Finally, we evaluated our solution's resilience to between 47 and 60 out-of-scope groups, which contributed 27\% to 87\% of the test samples.
\end{enumerate}

\subsection{Dataset preparation}

We required an initial malware dataset that had been used in prior work, contained both attribution metadata and malware samples, included multiple APT groups with varied sample frequencies, and was publicly available. We selected the APTMalware dataset~\cite{Cyber-research/APTMalware:Samples}, which contains samples from 12 APT groups with substantial class imbalance. The number of samples ranges from 32 samples, or 0.89\%, to 961 samples, or 26.74\% (see Table~\ref{tab:APTMalware dataset} in Appendix~\ref{sec:Sample Sets}). This imbalance is representative of operational settings and is also a known challenge for machine learning classifiers. At a later stage, we incorporated executable samples from the ADAPT Group-labelled Dataset~\cite{SecPriv/adapt:Samples} to perform large scale out-of-scope testing.

Before training, we extracted features from the malware samples and transformed them into numerical representations. APT groups may reuse tooling, code fragments, metadata conventions, and operational practices across campaigns. These recurring artefacts can provide attribution signals, although their strength varies across groups and samples.

Feature extraction was not the focus of this work, so we adopted the ADAPT feature extractor~\cite{Saha2024ADAPTFiles,SecPriv/adapt:Samples}, which uses a range of lightweight static analysis tools to extract features. We evaluated their FEAT, semantic string embedding, and combined feature sets, and found their semantic string embedding feature set to be the most effective for our dataset.

\subsection{Multi-class and binary classifiers}\label{sec:multi-class models}
We trained both multi-class and binary classifiers on the prepared feature representations. The dataset was split into training data (80\%) and test data (20\%) using a stratified split, ensuring that each APT group was represented proportionally in both sets.
The test set was held back until classifier selection and tuning were complete, and was used only to assess generalisation. For binary classification, labels were converted to `1' when the sample belonged to the APT group being modelled and `0' otherwise.
See Appendix~\ref{sec:model building} for more details.

\subsection{Ranked binary classifiers}
The binary classifiers were combined to perform selective multi-class attribution. 
We ranked the binary classifiers according to our confidence in their predictions, using validation performance and the number of training samples. The classifiers were then applied sequentially: a sample was assigned to the first classifier that accepted it above its tuned threshold. If no classifier accepted the sample, it remained unclassified through abstention.

This ranking strategy avoids directly selecting the classifier with the highest predicted probability, as in a standard One-vs-Rest (OVR) approach. OVR can be problematic when binary classifier quality varies across APT groups. For example, a poorly calibrated or overly permissive classifier could assign high probabilities to many samples and override a more reliable classifier.
This risk is particularly relevant because some APT groups have few training samples.
OVR would also require probability calibration, and even then, probabilities may not be directly comparable across different classifier types.

The ranked binary procedure is summarised below:
\begin{enumerate}
        \item Define the abstention and out-of-scope labels as additional labels beyond the groups included during training.
    \item If some classes were excluded during training, then:
    \begin{enumerate}
        \item Take a copy of the test labels;
        \item Replace labels for excluded groups with the out-of-scope label;
        \item Adjust the remaining labels to ensure a contiguous label range.
    \end{enumerate}
    \item Sort the precision classifiers, if to be used, according to their training precision, recall, and number of samples.  
    \item Sort the F1 classifiers, if to be used, according to their training precision, recall, and number of samples, and append them to the sorted precision classifiers.
    \item Initialise all test sample classifications as abstained.
    \item For each classifier in ranked order, using its tuned threshold, extract the list of samples that it classifies as originating from its APT group, and are still initialised as abstained, and allocate these samples to that classifier's APT group.
    \item     After all classifiers have been applied, leave any unassigned samples as abstained.
\end{enumerate}

\subsection{Out-of-scope group testing}

\begin{figure}[h!]
    \centering
    \includegraphics[width=1\linewidth]{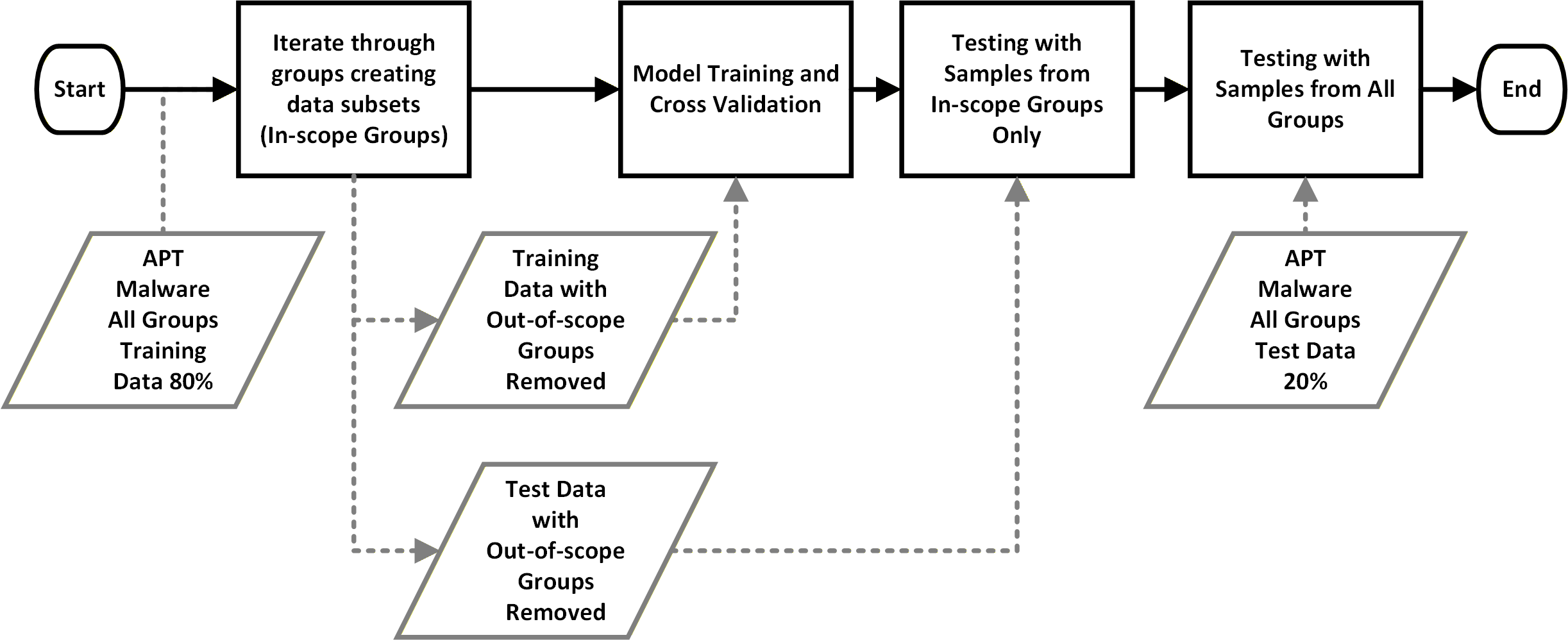}
    \caption{Experiments with out-of-scope groups}
    \label{fig:MC with OOS}
\end{figure}

To evaluate how the classifiers respond to samples from out-of-scope groups, we conducted a leave-one-group-out set of experiments, illustrated in Figure~\ref{fig:MC with OOS}.

In each experiment, one APT group was designated as out-of-scope. Its samples were removed from the training set and from the reduced test set used for post-validation testing. The full test set, including samples from the excluded group, was held back for the final out-of-scope evaluation. This process was repeated for each APT group, and the results were averaged.

For the baseline multi-class model, samples from the excluded group must be assigned to one of the known classes, since the model has no abstention mechanism. These experiments therefore provided a closed-set baseline and allowed us to analyse misclassification patterns.

For the ranked binary classifiers, it was not sufficient to remove only the binary classifier corresponding to the excluded group, since the remaining classifiers would still have seen that group's samples as negative examples during training. We therefore retrained the full set of binary classifiers after removing the excluded group's samples. The ranking logic was unchanged: out-of-scope samples were assigned to a known group only if accepted by one of the ranked binary classifiers; otherwise, they remained abstained.

\subsection{Large scale out-of-scope group testing}
\begin{figure}[h!]
    \centering
    \includegraphics[width=1\linewidth]{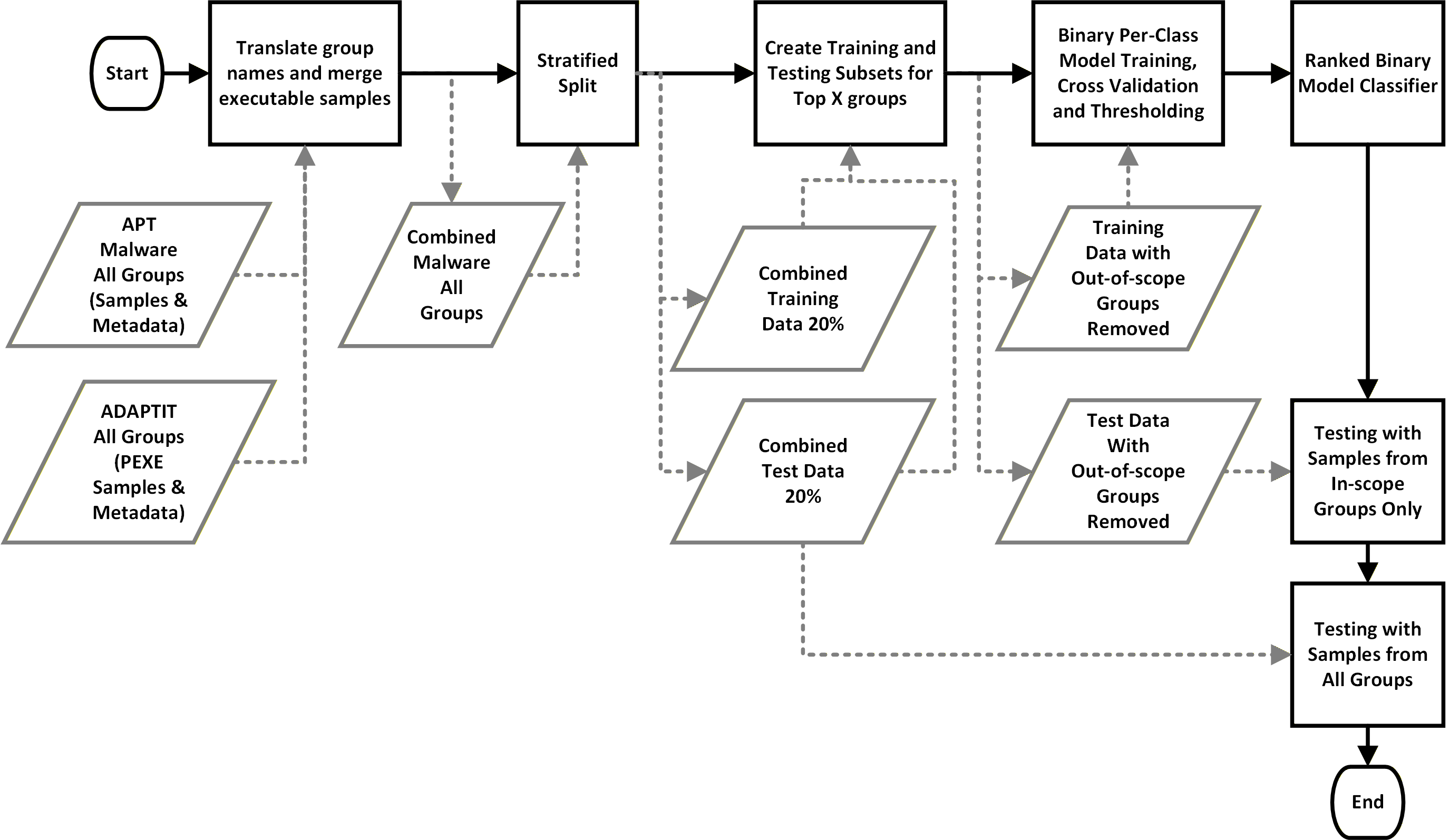}
    \caption{Large scale out-of-scope group testing}
    \label{fig:large scale OOS}
\end{figure}

To stress test the methodology's resilience in the presence of a large number of samples and groups we combined the executable samples from the ADAPT Group Attribution dataset~\cite{SecPriv/adapt:Samples} with the APT Malware dataset~\cite{CoenBoot2019ApplyingAttribution}. The end-to-end process is shown in Figure~\ref{fig:large scale OOS}.

Before merging the datasets, we mapped APT group names to a common namespace, since some groups appeared under different names in the two datasets. We also identified 514 overlapping samples, 99\% of which had matching labels. This supported the decision to combine the datasets by adding the new unique samples. We then rebuilt and tested the ranked binary classifiers on the combined dataset. Unlike the leave-one-group-out experiments, this evaluation excluded between 47 and 60 APT groups from training at the same time, allowing us to test resilience under large-scale out-of-scope exposure.

\section{Metrics}
\label{sec:metrics}

Let us now introduce the used metrics. For our experiments, three classes of metrics were required. 

\paragraph{Standard classification metrics}
The standard classification metrics that we used for binary classifiers were \emph{precision}, \emph{recall} and \emph{F1}.
Precision measures how often a binary classification is correct and uses the formula:
\[
\text{Precision} = \frac{\text{TP}}{\text{TP} + \text{FP}}
\]
where:
\begin{itemize}
    \item $\text{TP}$ = True Positives
    \item $\text{FP}$ = False Positives
\end{itemize}
Recall measures how often a binary class is correctly identified and uses the formula:
\[
\text{Recall} = \frac{\text{TP}}{\text{TP} + \text{FN}}
\]
where:
\begin{itemize}
    \item $\text{FN}$ = False Negatives
\end{itemize}F1 is the harmonic mean of precision and recall.

With highly unbalanced datasets, the accuracy metric is less meaningful, especially when building binary classifiers. For example, where we have APT groups that contribute only 1\% of the samples, a binary classifier that does not spot any samples would still be 99\% accurate. For this reason, we focus primarily on precision, recall and F1.

For multi-class models, precision, recall, and F1 can be computed per class and then averaged using class support as weights. We refer to these as \emph{weighted} metrics.
If samples of an APT group are rare in our datasets, then given how the datasets were created, they are also likely to be rare in practice.
Hence, we believe that weighted metrics best reflect the real world capabilities of our solutions and consider them as  key metrics.
For completeness, and to support comparison with prior work, we also report \emph{macro} metrics, where each class contributes equally regardless of the number of samples i.e.,
the metrics are calculated per class and the class scores then averaged.

\paragraph{Metric changes with abstentions}
When a model can abstain, we distinguish between the proportion of samples that are abstained or classified and the quality of the classifications made. The standard precision, recall, and F1 metrics remain useful for comparison with the closed-set baseline because they are calculated over the full dataset. We therefore retain them as comparative metrics, and introduce the following key metrics for selective classification:
\begin{itemize}
    \item \emph{Coverage}: the percentage of samples for which the model makes a classification.
    \item \emph{Selective Accuracy}: accuracy calculated only over classified samples.
    \item \emph{Selective F1}: F1 score calculated only over classified samples.
\end{itemize}
There is no need to add selective precision metrics as they are equivalent to standard precision.
Recall would normally be amended to penalise abstentions as follows:
\[
\text{Effective Recall} = \frac{\text{TP}}{\text{TP} + \text{FN} + \text{Abstentions}}
\]
However, because we treat abstentions as a separate class, the standard recall metric will automatically count abstentions as false negatives and 
therefore capture the effect of abstention without requiring a separate effective recall metric.

\paragraph{Metric changes with out-of-scope data}
We now include the \emph{out-of-scope rejection rate} metric which measures the percentage of out-of-scope samples that are correctly abstained. Coverage is also constrained to \emph{in-scope coverage} as coverage is now only desirable for samples from groups represented during training. 
Recall continues to be calculated over samples that have the potential to be correctly identified, i.e., in-scope samples. 

\paragraph{Metric Abbreviations}
In the following tables, we use the following abbreviations to save space: precision (Prec), recall (Rec), macro (m), coverage (Cov), selective (S-), out-of-scope rejection rate (OOS-RR), in-scope (is).
A conceptual summary of the metrics used throughout the experiments can be found in Table \ref{tab:standard metrics}, Appendix \ref{sec:Conceptual Metrics Descriptions}.

\begin{table*}[!t]
\caption{Multi-class models}\label{tab:multi-class-classifiers}
\centering
\begin{tabular}{ll|rrr|rrr}
\hline
      &        & \multicolumn{3}{c|}{Key Metrics} & \multicolumn{3}{c}{Comparative Metrics} \\
Model & Tuning & Prec & Rec & F1 & Prec(m) & Rec(m) & F1(m) \\
\hline
Cat Boost & F1 & 90\% & 89\% & 89\% & 89\% & 85\% & 86\% \\
Extra Trees & F1 & 90\% & \textbf{90}\% & \textbf{90}\% & \textbf{92}\% & 85\% & \textbf{90}\% \\
Hist Gradient Boosting & F1 & 90\% & \textbf{90}\% & \textbf{90}\% & 89\% & 85\% & 86\% \\
LGBM & F1 & 90\% & \textbf{90}\% & 89\% & 91\% & 83\% & 86\% \\
Random Forest & F1 & 78\% & 74\% & 75\% & 69\% & 76\% & 69\% \\
XGB & F1 & 90\% & \textbf{90}\% & \textbf{90}\% & 91\% & 83\% & 89\% \\
\hline
Cat Boost & Precision & 90\% & 89\% & 89\% & 89\% & 85\% & 86\% \\
Extra Trees & Precision & 90\% & \textbf{90}\% & \textbf{90}\% & \textbf{92}\% & 85\% & 88\% \\
\textbf{Hist Gradient Boosting} & \textbf{Precision} & \textbf{91}\% & \textbf{90}\% & \textbf{90}\% & 91\% & \textbf{87}\% & 89\% \\
LGBM & Precision & 90\% & \textbf{90}\% & \textbf{90}\% & 91\% & 83\% & 86\% \\
Random Forest & Precision & 78\% & 74\% & 75\% & 69\% & 76\% & 69\% \\
XGB & Precision & 90\% & \textbf{90}\% & \textbf{90}\% & 91\% & 83\% & 86\% \\
\hline
Xu \cite{Xu2021AnLightGBM} & &&&& 90\% & 85\% & 87\% \\
Kida \cite{Kida2023Nation-StateHashing} & &&&&&& 88\% \\
\hline
\end{tabular}
\end{table*}

\begin{table*}[h!]
\caption{Per-APT group level comparison of binary classifiers vs. the baseline multi-class model} 
\label{tab:binary-classifiers-v-best-mc}
\centering
\begin{tabular}{ll|lrrr|rrr}
\hline
    &  & \multicolumn{4}{c|}{Binary Classifiers} & \multicolumn{3}{c}{Baseline Multi-class} \\
APT Group & Tuning & Best Classifier & Prec & Rec & F1 & Prec & Rec & F1 \\
\hline
APT 1          & F1        & Hist Gradient Boosting &  97\% & 89\% & \textbf{93}\% & 90\% & 93\% & 91\% \\
APT 10         & F1        & Hist Gradient Boosting &  97\% & 69\% & \textbf{81}\% & 89\% & 69\% & 78\% \\
APT 19         & F1        & Extra Trees            & 100\% & 57\% & \textbf{73}\% & 75\% & 43\% & 55\% \\
APT 21         & F1        & Hist Gradient Boosting &  95\% & 86\% & 90\% & 100\% & 86\% & \textbf{93}\% \\
APT 28         & F1        & Hist Gradient Boosting &  89\% & 56\% & 69\% & 83\% & 67\% & \textbf{74}\% \\
APT 29         & F1        & Hist Gradient Boosting &  80\% & 86\% & \textbf{83}\% & 72\% & 86\% & 78\% \\
APT 30         & F1        & Hist Gradient Boosting &  88\% & 88\% & 88\% & 88\% & 91\% & \textbf{90}\% \\
Dark Hotel     & F1        & XGB                    &  94\% & 80\% & \textbf{86}\% & 84\% & 85\% & 85\% \\
Energetic Bear & F1        & Extra Trees            & 100\% & 93\% & 96\% & 100\% & 93\% & 96\% \\
Equation Group & F1        & Hist Gradient Boosting & 100\% & 99\% & 99\% & 100\% & 99\% & 99\% \\
Gorgon Group   & F1        & Hist Gradient Boosting &  94\% & 96\% & \textbf{95}\% & 90\% & 96\% & 93\% \\
Winti          & F1        & Hist Gradient Boosting &  99\% & 94\% & \textbf{96}\% & 94\% & 95\% & 94\% \\
\hline
\textbf{Mean} & F1 &  & 90\% & 83\% & \textbf{87}\% & 89\% & 84\% & 86\% \\
\hline
APT 1          & Precision & Hist Gradient Boosting & \textbf{100}\% & 78\% & 88\% & 90\% & 93\% & 91\% \\
APT 10         & Precision & Extra Trees            &  \textbf{96}\% & 55\% & 70\% & 89\% & 69\% & 78\% \\
APT 19         & Precision & Hist Gradient Boosting & 100\% & 71\% & 83\% & 100\% & 71\% & 83\% \\
APT 21         & Precision & Extra Trees            & 100\% & 77\% & 87\% & 100\% & 86\% & 93\% \\
APT 28         & Precision & Hist Gradient Boosting & \textbf{100}\% & 40\% & 57\% & 82\% & 74\% & 78\% \\
APT 29         & Precision & Hist Gradient Boosting &  \textbf{90}\% & 67\% & 77\% & 75\% & 86\% & 80\% \\
APT 30         & Precision & Hist Gradient Boosting &  \textbf{94}\% & 88\% & 91\% & 88\% & 91\% & 90\% \\
Dark Hotel     & Precision & Hist Gradient Boosting &  \textbf{97}\% & 65\% & 78\% & 84\% & 85\% & 85\% \\
Energetic Bear & Precision & Extra Trees            & 100\% & 93\% & 96\% & 100\% & 93\% & 96\% \\
Equation Group & Precision & Hist Gradient Boosting & 100\% & 99\% & 99\% & 100\% & 99\% & 99\% \\
Gorgon Group   & Precision & Hist Gradient Boosting &  \textbf{99}\% & 84\% & 91\% & 91\% & 96\% & 96\% \\
Winti          & Precision & Extra Trees            & \textbf{100}\% & 86\% & 92\% & 96\% & 96\% & 96\% \\
\hline
\textbf{Mean} & Precision &  & \textbf{98}\% & 75\% & 84\% & 91\% & 87\% & 89\% \\
\hline
\textbf{Combined Mean} &  &  & \textbf{96}\% & 79\% & 86\% & 90\% & 85\% & \textbf{87}\% \\
\hline
\end{tabular}
\end{table*}

\section{Experimental results}
\label{sec:results}
In this section, we introduce the results of our experiments.
In the first set of experiments, we created a baseline by tuning a variety of malware based, multi-class, APT group classification models. This resulted in a best-of-class classifier that we refer to as the \emph{baseline multi-class model}.
To allow APT group specific tuning, we then created two binary APT group classifiers for each APT group. We then compared them group-for-group to the baseline multi-class model to ensure we had improved F1 and/or precision.
In order to produce an enhanced classifier, we then ranked the binary classifiers and combined them to simulate a new multi-class classifier, and compared it to the baseline multi-class model. Finally, to test how our solution would work in an operational environment, we carried out sets of experiments to test how well both the baseline multi-class model and our ranked binary classifiers performed in the presence of out-of-scope samples, i.e., malware samples from groups they were not trained on.

\subsection{Multi-class and binary classifiers}
This subsection establishes the baseline multi-class model and evaluates whether APT group specific binary classifiers improve per-group precision or F1 score.
The results when building a range of multi-class classifiers over the complete set of APT groups can be seen in Table~\ref{tab:multi-class-classifiers}. The differences in the overall metrics for most models were not significant, except for Random Forest which was far lower on all metrics.
The overall best multi-class model was the \emph{Hist Gradient Boosting} model tuned for precision.
This outperformed existing models in the literature built against the same dataset \cite{Xu2021AnLightGBM, Kida2023Nation-StateHashing}. We used the \emph{Hist Gradient Boosting} model tuned for precision as our baseline to experiment with other methods and is referred to in the rest of this paper as the \emph{baseline multi-class model}.

We built a pair of binary classifiers for each APT group, compared them at the APT group level with the baseline multi-class model, and report the results in Table \ref{tab:binary-classifiers-v-best-mc}.
As the APT Malware dataset has 12 APT groups (see Table~\ref{tab:APTMalware dataset}, Appendix~\ref{sec:Sample Sets}) this resulted in 12 binary F1-tuned and 12 binary precision-tuned classifiers. 
The numbers in the last three columns were derived by taking the baseline multi-class model from Table~\ref{tab:multi-class-classifiers} and calculating its metrics for each APT group. 

The results of Table~\ref{tab:binary-classifiers-v-best-mc} indicate that different classifier types provided better results for different APT groups and tunings, with Hist Gradient Boosting selected the most, followed by Extra Trees. For models tuned for F1 score, the binary classifiers typically outperformed the baseline multi-class model in terms of per-APT group F1. Likewise, for models tuned for precision, the binary classifiers typically achieved higher per-APT group precision than the baseline multi-class model.

Overall, the binary classifiers achieved an average precision of 96\%, recall of 79\%, and F1 score of 86\%, compared with 90\%, 85\%, and 87\%, respectively, for the baseline multi-class model. 
These results show that the binary modelling approach improves precision, but at the cost of reduced recall, resulting in an overall F1 score that remains comparable to the baseline multi-class model.

\subsection{Ranked binary classifiers}

\begin{table*}[h!]
\caption{Ranked binary classifiers compared with multi-class models over the same dataset}
\label{tab:combined-binary-classifiers}
\centering
\begin{tabular}{l|lll|llllll}
\hline
& \multicolumn{3}{l|}{Key Metrics} & \multicolumn{6}{l}{Comparative Metrics}\\
& Cov & S-Acc & S-F1 & Prec & Rec & F1 & Prec(m) & Rec(m) & F1(m) \\
\hline
Ranked Binary F1 &             \textbf{92}\% & 95\% & 95\% & 96\% & 87\% & 91\% & 96\% & 75\% & 80\% \\
Ranked Binary Precision &      79\% & \textbf{98}\% & \textbf{98}\% & \textbf{98}\% & 78\% & 87\% & \textbf{98}\% & 69\% & 77\% \\
Ranked Binary Precision + F1 & \textbf{92}\% & 95\% & 95\% & 95\% & 88\% & 91\% & 96\% & 77\% & 82\% \\
\hline
Baseline Multi-class Model &  &  &  & 91\% & \textbf{90}\% & 90\% & 91\% & \textbf{87}\% & \textbf{89}\% \\
\hline
Xu \cite{Xu2021AnLightGBM} & & & &&&& 90\% & 85\% & 87\% \\
Kida \cite{Kida2023Nation-StateHashing} &  &  &  & & & & && 88\% \\
\hline
\end{tabular}
\end{table*}

Table~\ref{tab:combined-binary-classifiers} reports the results of ranked binary classifiers to simulate multi-class classification. We compare these results against both previously published results on the same dataset, and our baseline multi-class model.
We evaluate three ranking strategies: binary classifiers tuned for F1 score, binary classifiers tuned for precision, and the hybrid strategy in which precision-tuned classifiers are applied before F1-tuned classifiers. We refer to the latter configuration as \emph{precision + F1}.
Across all ranked binary configurations, precision increased substantially compared to the baseline multi-class model. This improvement was achieved while either slightly increasing F1, or reducing it by only 3\%. This shift towards precision, with a balanced reduction in recall improves the practical applicability of the approach in real-world settings where false positives carry operational costs. 
Because prior work on this dataset does not report weighted metrics, direct comparison with published results is limited to the macro metrics. Under this comparison, our approach substantially improves macro precision, but with corresponding reductions in macro recall and macro F1. Although our overall weighted F1 performance remains high, a small number of APT groups exhibit weaker recall. These lower per-group scores have a stronger effect on the macro-averaged metrics, which weight each group equally regardless of its sample size.

\begin{figure*}[!t]
\centering
\subfloat[Baseline multi-class model]{\includegraphics[width=3in]{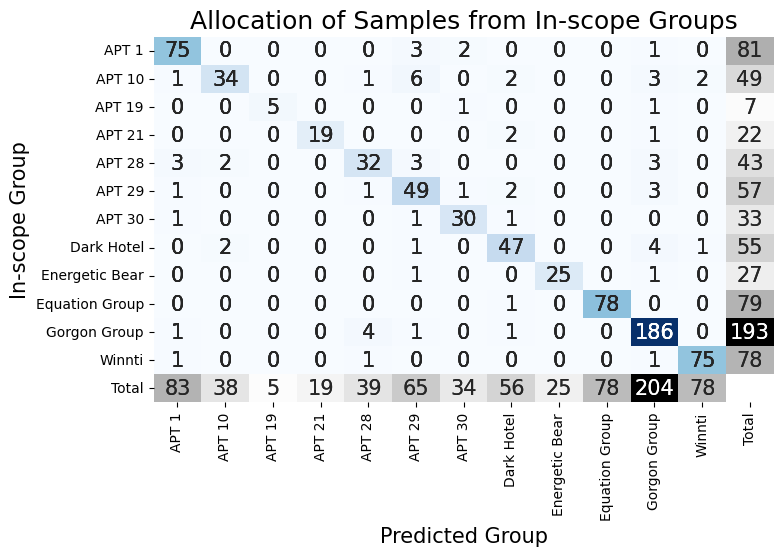}
\label{fig:Hist Gradient Boost allocation of in-scope}}
\hfil
\subfloat[Ranked binary precision classifiers]{\includegraphics[width=3.2in]{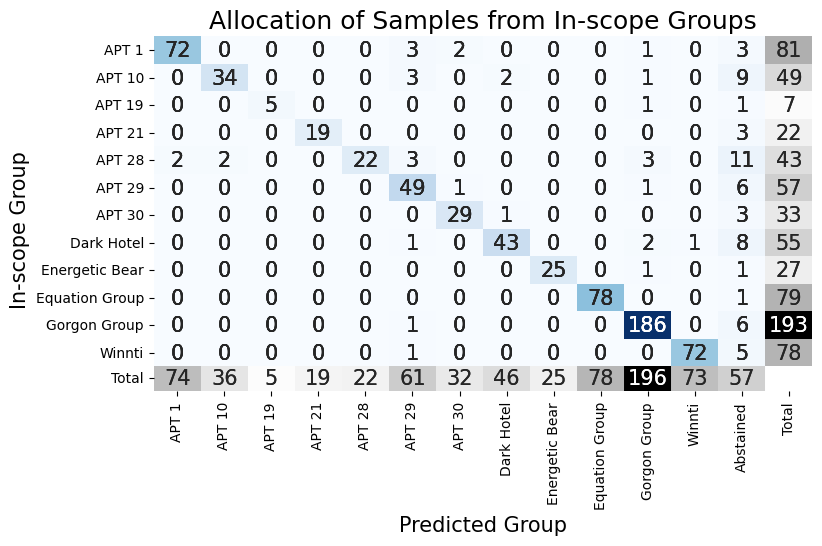}%
\label{fig:Precision Models allocation of in-scope}}
\caption{Distribution of in-scope samples with different approaches}
\label{fig:in-scope}
\end{figure*}

Figure~\ref{fig:in-scope} shows how in-scope samples are classified by the baseline multi-class model and the ranked binary \emph{precision + F1} classifiers. The vertical axis represents the true APT group, while the horizontal axis represents the predicted classification.
Through this visualisation, we observed a clear difference in how samples that could not be reliably classified were treated. 
The baseline multi-class model is forced to assign every sample to one of the known APT groups, leading to misclassifications across several classes, as shown in Figure~\ref{fig:Hist Gradient Boost allocation of in-scope}. By contrast, the ranked binary \emph{precision + F1} classifiers often abstain from classification when confidence is insufficient, as shown in Figure~\ref{fig:Precision Models allocation of in-scope} column 13. This abstention mechanism reduces incorrect attributions and contributes to the improved precision observed in the ranked binary setting.

\begin{table}[h!]
\caption{Classifications relative to the baseline multi-class model}
\label{tab:effect on classifications}
\centering
\begin{tabular}{l|rr}
\hline
Classifiers & Incorrect & Correct \\
\hline
Ranked binary F1             & -49\% &  -4\% \\
Ranked binary Precision      & -87\% & -14\% \\
Ranked binary precision + F1 & -52\% &  -3\% \\
\hline
\end{tabular}
\end{table}

We introduce in Table~\ref{tab:effect on classifications} the percentage change in the number of correct and incorrect classifications made when switching from the baseline multi-class model to the ranked binary classifiers.
The ranked binary \emph{precision + F1} classifiers reduce misclassifications by 52\%, while removing only 3\% of the correct classifications. The ranked binary precision classifiers are more selective, achieving an 87\% reduction in misclassifications, at the cost of a 14\% reduction in correct classifications. The ranked binary F1 classifiers offer the weakest reduction in incorrect classifications and removes more correct classifications than the \emph{precision + F1} approach. Consequently, it was excluded from the subsequent experiments.

\subsection{Out-of-scope group testing}

\begin{table*}[h!]
\caption{Baseline multi-class model with out-of-scope groups (no abstention)}
\label{tab:HGB-P-with-unknowns}
\centering
\begin{tabular}{l|rrr|rrr}
\hline
& \multicolumn{3}{c|}{Key Metrics} & \multicolumn{3}{c}{Comparative Metrics} \\
Out-of-scope Group & Prec & Rec & F1 & Prec(m) & Rec(m) & F1(m)  \\
\hline
None (baseline with all groups in-scope) & 91\% & 90\% & 90\% & 91\% & 87\% & 89\% \\
\hline
APT 1           & 72\% & 89\% & 75\% & 73\% & 84\% & 74\% \\
APT 10          & 80\% & 91\% & 82\% & 80\% & 88\% & 80\% \\
APT 19          & 89\% & 90\% & 89\% & 82\% & 88\% & 81\% \\
APT 21          & 85\% & 90\% & 86\% & 80\% & 85\% & 78\% \\
APT 28          & 82\% & 92\% & 84\% & 81\% & 87\% & 80\% \\
APT 29          & 78\% & 91\% & 80\% & 77\% & 86\% & 77\% \\
APT 30          & 82\% & 90\% & 84\% & 80\% & 84\% & 78\% \\
Dark Hotel      & 78\% & 90\% & 80\% & 77\% & 86\% & 78\% \\
Energetic Bear  & 85\% & 90\% & 85\% & 79\% & 85\% & 78\% \\
Equation Group  & 76\% & 89\% & 76\% & 79\% & 84\% & 76\% \\
Gorgon Group    & 54\% & 88\% & 58\% & 70\% & 84\% & 70\% \\
Winti           & 72\% & 89\% & 75\% & 72\% & 85\% & 74\% \\
\hline
Mean            & 78\% & 90\% & 80\% & 78\% & 86\% & 77\% \\
Delta from baseline &-13\% &  0\% &-10\% &-13\% & -1\% &-12\% \\ 
\hline
\end{tabular}
\end{table*}

We evaluated the resilience of the baseline multi-class model to samples from out-of-scope APT groups. 
In each experiment, one group was designated as the out-of-scope group: its samples were removed from the training set but retained in the final test set. This procedure was repeated for all twelve groups and the results averaged across the resulting experiments. 
The results are shown in Table~\ref{tab:HGB-P-with-unknowns}. As expected, precision decreased when out-of-scope samples were included. Since the baseline multi-class model operates in a closed-set setting, it must assign every input to one of the known training classes. As a result, samples from unseen groups are necessarily assigned to an incorrect known class. This limitation reduces the practical applicability of closed-set multi-class classifiers in operational environments, where previously unseen groups may be encountered.

\begin{figure*}[!t]
\centering
\subfloat[Baseline multi-class model]{\includegraphics[width=3in]{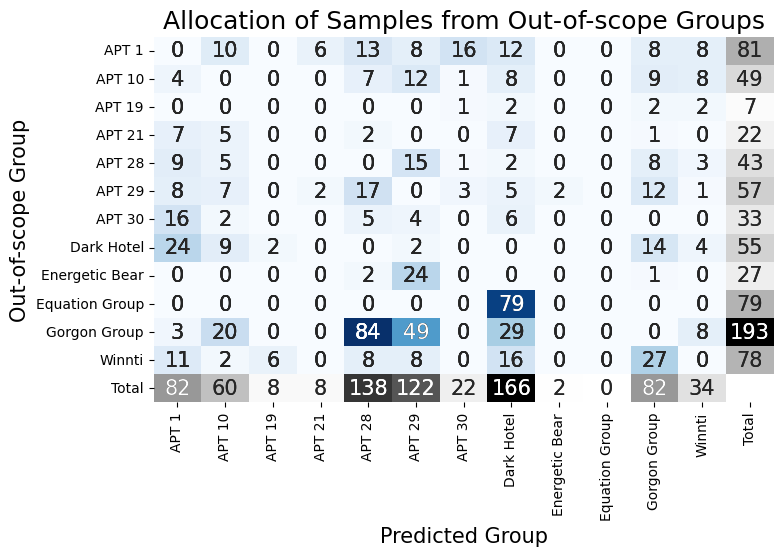}%
\label{fig:Hist Gradient Boost allocation of out-of-scope}}
\hfil
\subfloat[Ranked binary precision classifiers]{\includegraphics[width=3.2in]{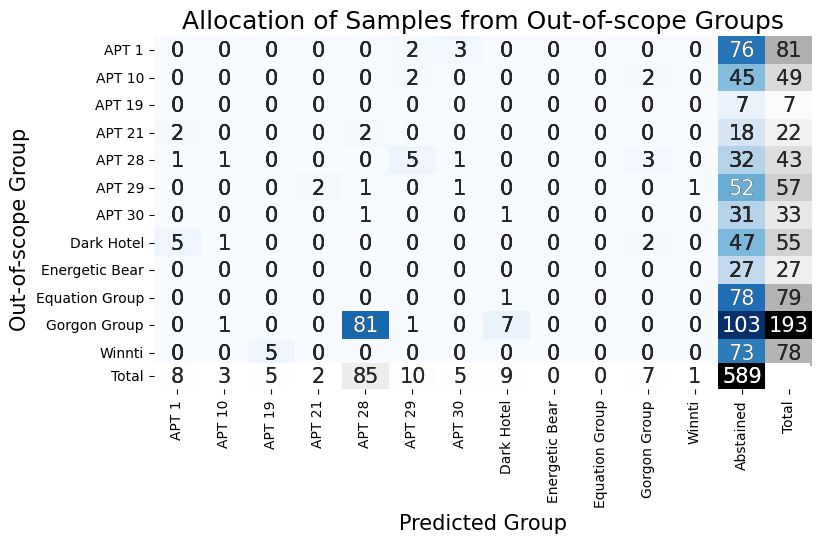}%
\label{fig:Precision Models allocation of out-of-scope}}
\caption{Distribution of out-of-scope samples with different approaches}
\label{fig:OOS}
\end{figure*}

Figure~\ref{fig:OOS} compares how out-of-scope samples are classified by the baseline multi-class model and the ranked binary precision classifiers. Each row represents a separate leave-one-group-out experiment, where the corresponding APT group was removed from the training set but retained during testing. The vertical axis indicates the excluded APT group, while the horizontal axis indicates the class assigned by the model when evaluated on the full dataset.
Figure~\ref{fig:Hist Gradient Boost allocation of out-of-scope} illustrates  the behaviour of the baseline multi-class model in this setting, and how each of the APT group's samples were 
misclassified when excluded from training.
For example,  when the Gorgon Group was excluded from training, many of its 193 samples were misclassified as APT28 or APT29, as shown in row 11 in Figure~\ref{fig:Hist Gradient Boost allocation of out-of-scope}. This type of error is particularly problematic from an operational perspective: MITRE ATT\&CK \cite{MITREATTCK} describes the Gorgon Group as suspected to be Pakistan-based, whereas APT28 and APT29 are suspected to be Russia-based.  The nature of these misclassifications is significant, as  the model does not merely make an incorrect classification, but may also imply a substantially different attribution context. Misclassifications into unrelated groups can have a greater practical impact than errors between groups with similar behaviours, tooling, or defensive countermeasures.

\begin{table*}[h!]
\caption{Ranked binary precision + F1 classifiers with out-of-scope groups}
\label{tab:PF1-with-unknowns}
\centering
\begin{tabular}{l|rrrr|rrrrrr}
\hline
Out-of-scope & \multicolumn{4}{l|}{Key Metrics} & \multicolumn{3}{l}{Comparative Metrics} \\
Group & Cov(is) & OOS-RR & S-Acc & S-F1 & Prec & Rec(is) & F1 & Prec(m) & mRec(is) & F1(m) \\
\hline
None (baseline with all groups in-scope)  & 92\% &      & 95\% & 95\% & 95\% & 88\% & 91\% & 96\% & 77\% & 82\% \\
\hline
APT 1           & 93\% & 54\% & 89\% & 87\% & 85\% & 88\% & 86\% & 81\% & 80\% & 82\% \\
APT 10          & 94\% & 43\% & 90\% & 88\% & 87\% & 89\% & 88\% & 84\% & 80\% & 82\% \\
APT 19          & 92\% & 43\% & 94\% & 94\% & 94\% & 88\% & 91\% & 87\% & 81\% & 82\% \\
APT 21          & 92\% & 41\% & 93\% & 92\% & 91\% & 87\% & 89\% & 86\% & 80\% & 82\% \\
APT 28          & 93\% & 44\% & 92\% & 91\% & 90\% & 89\% & 89\% & 85\% & 83\% & 85\% \\
APT 29          & 92\% & 63\% & 93\% & 91\% & 90\% & 88\% & 89\% & 86\% & 81\% & 84\%\\
APT 30          & 93\% & 64\% & 93\% & 92\% & 92\% & 89\% & 90\% & 84\% & 83\% & 84\% \\
Dark Hotel      & 93\% & 38\% & 90\% & 88\% & 86\% & 88\% & 87\% & 82\% & 80\% & 81\% \\
Energetic Bear  & 93\% & 52\% & 92\% & 91\% & 91\% & 87\% & 89\% & 85\% & 78\% & 80\% \\
Equation Group  & 91\% &  0\% & 83\% & 80\% & 79\% & 86\% & 82\% & 82\% & 75\% & 77\% \\
Gorgon Group    & 91\% & 34\% & 74\% & 69\% & 67\% & 85\% & 75\% & 75\% & 79\% & 78\% \\
Winti           & 93\% & 68\% & 91\% & 89\% & 88\% & 87\% & 87\% & 80\% & 82\% & 82\% \\
\hline
Mean      & 93\% & 45\% & 90\% & 88\% & 87\% & 88\% & 87\% & 83\% & 80\% & 82\% \\
\hline
Delta from baseline  & +1\% &      & -5\% & -7\% & -8\% &  0\% & -4\% &-13\% & -3\% &  0\% \\
\hline
\end{tabular}
\end{table*}

After evaluating the baseline multi-class model under out-of-scope conditions, we repeated the same leave-one-group-out experiments using two ranked binary configurations: the ranked binary \emph{precision + F1} classifiers and the ranked binary precision classifiers. In the ideal case, all samples from the excluded APT group would be rejected by the binary classifiers and remain unclassified through abstention.
Table~\ref{tab:PF1-with-unknowns} presents the results for the ranked binary \emph{precision + F1} configuration across the twelve out-of-scope experiments. Compared to the baseline multi-class model under the same conditions, this approach increases precision by 9\%, reduces recall by only 2\%, and improves F1 score by 5\%. However, it abstains on only 45\% of out-of-scope samples. Thus, while the \emph{precision + F1} configuration improves robustness compared with closed-set multi-class classification, it still misassigns many unseen samples to known APT groups. 

Table~\ref{tab:P-with-unknowns} presents the results for the ranked binary precision configuration. This approach provides a much stronger abstention capability, rejecting 88\% of out-of-scope samples compared with 45\% for the ranked binary \emph{precision + F1} classifier. This makes the ranked binary precision classifiers substantially more suitable for operational settings in which samples may originate from APT groups not represented during training.
Figure~\ref{fig:Precision Models allocation of out-of-scope} shows the distribution of out-of-scope samples under the ranked binary precision classifiers. Compared with the baseline multi-class model in Figure~\ref{fig:Hist Gradient Boost allocation of out-of-scope}, the improvement is clear: most out-of-scope samples are rejected rather than being forced into one of the known classes.

The presence of out-of-scope samples has limited impact on most metrics when compared with the in-scope baseline, with most changes remaining within 2\%. The main exceptions are macro precision, which decreases by 8\%, and macro in-scope recall, which decreases by 5\%. This indicates that the classifier is less able to identify samples from some in-scope classes under out-of-scope conditions. However, because in-scope coverage decreases by only 1\% and selective accuracy increases by 1\%, the total percentage of samples that were correctly identified remains roughly unchanged.
The main outlier is the Gorgon Group: when treated as out-of-scope, many of its samples are still classified as APT28, as shown in Figure~\ref{fig:Precision Models allocation of out-of-scope}. This suggests that, although the ranked binary precision classifier substantially improves abstention overall, some unseen groups remain difficult to separate from specific known classes.

\begin{table*}[h!]
\caption{Ranked binary precision classifiers with out-of-scope groups}
\label{tab:P-with-unknowns}
\centering
\begin{tabular}{l|rrrr|rrrrrr}
\hline
Out-of-scope & \multicolumn{4}{l|}{Key Metrics} & \multicolumn{3}{l}{Comparative Metrics} \\
Group & Cov(is) & OOS-RR & S-Acc & S-F1 & Prec & Rec(is) & F1 & Prec(m) & mRec(is) & F1(m) \\
\hline
None (baseline with all groups in-scope)& 79\% &      & 98\% & 98\% & 98\% & 78\% & 87\% & 98\% & 69\% & 77\% \\
\hline
APT 1           & 79\% & 94\% & 98\% & 97\% & 97\% & 78\% & 86\% & 89\% & 74\% & 78\% \\
APT 10          & 79\% & 92\% & 99\% & 98\% & 98\% & 79\% & 87\% & 90\% & 74\% & 79\% \\
APT 19          & 79\% &100\% & 99\% & 99\% & 99\% & 78\% & 87\% & 98\% & 76\% & 77\% \\
APT 21          & 74\% & 82\% & 98\% & 97\% & 97\% & 73\% & 83\% & 89\% & 71\% & 75\% \\
APT 28          & 82\% & 74\% & 96\% & 96\% & 95\% & 81\% & 87\% & 88\% & 80\% & 82\% \\
APT 29          & 75\% & 91\% & 99\% & 98\% & 98\% & 74\% & 84\% & 90\% & 74\% & 78\% \\
APT 30          & 75\% & 94\% & 99\% & 99\% & 99\% & 74\% & 85\% & 90\% & 73\% & 77\% \\
Dark Hotel      & 82\% & 85\% & 97\% & 96\% & 96\% & 81\% & 88\% & 89\% & 77\% & 80\% \\
Energetic Bear  & 77\% &100\% & 99\% & 99\% & 99\% & 76\% & 86\% & 99\% & 72\% & 75\% \\
Equation Group  & 77\% & 99\% & 98\% & 98\% & 98\% & 76\% & 86\% & 90\% & 74\% & 79\% \\
Gorgon Group    & 78\% & 53\% & 81\% & 77\% & 76\% & 77\% & 76\% & 81\% & 73\% & 77\% \\
Winti           & 78\% & 94\% & 98\% & 98\% & 98\% & 77\% & 86\% & 77\% & 75\% & 79\% \\
\hline
Mean      & 78\% & 88\% & 97\% & 96\% & 96\% & 77\% & 85\% & 90\% & 74\% & 78\% \\
\hline
Delta from baseline  & -1\% &      & +1\% & -2\%  & -2\% & -1\% & -2\% & -8\% & -5\% & +1\% \\
\hline
\end{tabular}
\end{table*}

\subsection{Large scale out-of-scope group testing}
In operational environments, classifiers may encounter samples from many APT groups that were not represented during training. This may occur either due to a lack of sufficient samples, or the emergence of new groups.
In order to evaluate our methodology under these conditions, we tested the classifiers in the presence of large numbers of samples from previously unseen APT groups. We constructed this setting by merging the PEXE samples from the ADAPT Group-labelled Dataset~\cite{SecPriv/adapt:Samples} with the APT Malware dataset~\cite{Cyber-research/APTMalware:Samples}. We report the results from three of those experiments.

Table \ref{tab:combined} presents the results when training our classifiers over the top 12 groups in the combined dataset. Here 87\% of the test samples came from the 60 out-of-scope APT groups. This was our most extreme test of out-of-scope sample handling.
In this experiment, the ranked binary precision classifiers abstained from classifying 94\% of samples from out-of-scope groups, while achieving 92\% precision on classified samples, 95\% selective accuracy and covering 70\% of samples from in-scope groups. These results demonstrate strong resilience to out-of-scope samples under a highly imbalanced and operationally realistic setting. 
Details of the training data are provided in Table \ref{tab:Top12CombinedGroups}, Appendix \ref{sec:Sample Sets}.

Table \ref{tab:combined} also reports  the results for ranked binary precision classifiers trained on the top 20, and top 25 APT groups. In these settings, 42\% and 27\% of the test samples were from 53 and 47 out-of-scope groups, respectively.
In these experiments, ranked binary precision classifiers correctly abstained from classifying 92\% and 94\% of the samples from the out-of-scope groups, while achieving 93\% and 95\% precision on classified samples and  95\% and 96\% selective accuracy.  The classifiers also retained coverage over 72\% and 70\% of samples from in-scope groups, respectively. 
These results indicate that the ranked binary precision approach remains effective across different levels of out-of-scope exposure, maintaining high precision while rejecting the majority of unseen-group samples. Details of the corresponding training data are provided in Tables~\ref{tab:Top20CombinedGroups} and~\ref{tab:Top25CombinedGroups}, Appendix~\ref{sec:Sample Sets}.

\begin{table*}[h!]
\caption{Ranked binary precision classifiers on combined dataset of 72 groups}
\label{tab:combined}
\centering
\begin{tabular}{l|r|rrrr|rrrrrr}
\hline
& & \multicolumn{4}{l|}{Key Metrics} & \multicolumn{3}{l}{Comparative Metrics} \\
In-Scope & OOS & Cov(is) & OOS-RR & S-Acc & S-F1 & Prec & Rec(is) & F1 & Prec(m) & mRec(is) & F1(m) \\
\hline
Top 12 of 72 & 87\% & 70\% & 94\% & 95\% & 94\% & 92\% & 69\% & 79\% & 87\% & 68\% & 76\% \\
Top 20 of 72 & 42\% & 72\% & 92\% & 95\% & 94\% & 93\% & 71\% & 80\% & 89\% & 72\% & 79\% \\
Top 25 of 72 & 27\% & 70\% & 94\% & 96\% & 96\% & 95\% & 68\% & 80\% & 87\% & 61\% & 71\% \\
\hline
\end{tabular}
\end{table*}

\section{Discussion}
\label{sec:discussion}
\subsection{Implications of the Experimental Results}

To establish a baseline for evaluating our ranking binary approach,  we first tuned a set of multi-class models. This was necessary as prior works using the same dataset did not perform the full set of evaluations required for our analysis, nor did it report all of the metrics needed for comparison. Our best multi-class model outperformed previously published results on the same dataset~\cite{Xu2021AnLightGBM,Kida2023Nation-StateHashingb}, and was therefore used as our baseline multi-class model for subsequent experiments.

To allow APT group specific tuning, we created individual binary classifiers, one per APT group.
For each group, we dynamically selected the type of classifier and its hyper-parameters, and tuned its thresholds to achieve the maximum F1 and/or precision. 
In a binary classifier, tuning for F1 balances classification coverage with classification correctness. Whereas, tuning for precision prioritizes accurate classifications, but at the potential cost of producing fewer positive classifications (i.e., abstaining on more samples).
When tuning for F1 or precision, our binary classifiers had equal or higher F1 or precision values when compared per APT group with our baseline multi-class model.

To improve on our baseline multi-class model, we ranked and combined our binary classifiers to produce an enhanced multi-class attribution system. By switching from our baseline multi-class model to our ranked binary \emph{precision + F1} classifiers, we reduced incorrect classifications by 52\% while only reducing correct classifications by 3\%. We also showed a more extreme case where the use of the ranked binary precision classifiers reduced incorrect classifications by 87\%, but with a cost to correct classifications of 14\%. The \emph{precision + F1} configuration provides broader coverage while substantially reducing errors, whereas the ranked precision configuration produces fewer classifications but offers higher confidence in the classifications it does make. 
The appropriate operating point will depend on the deployment context: broader coverage may be preferable when attribution is used to support exploratory analysis, while higher precision may be preferable when classifications directly inform defensive action.

This trade-off is particularly important in operational settings, where samples from APT groups not represented during training are likely to occur. We first confirmed experimentally that the baseline multi-class model necessarily misclassifies samples from out-of-scope groups, since it is forced to assign every sample to one of the known classes. Under the same conditions, when 8\% of samples came from out-of-scope groups, the ranked binary \emph{precision + F1} classifiers abstained on 45\% of out-of-scope samples, while the ranked precision classifiers abstained on 88\%. In both cases, the effect on most other metrics was limited relative to the corresponding in-scope baseline. This demonstrates the value of explicit abstention: rather than producing unsupported attributions for unseen groups, the classifiers can withhold classification when the available evidence is insufficient.

The ranked binary precision configuration is therefore the better fit for real-world use when the cost of incorrect attribution is high. Although this configuration reduces recall, the reduction is offset by a substantial decrease in incorrect classifications and a much stronger ability to reject out-of-scope samples. In security operations, this trade-off is often desirable: a smaller number of high-confidence classifications may be more actionable than broader coverage accompanied by a higher risk of misleading attribution.

Finally, to simulate a harsher operational environment, we evaluated the ranked precision classifier in the presence of samples from 60 out-of-scope groups, which together accounted for 87\% of the test data. In this extreme setting, our solution abstained on 94\% of out-of-scope samples. For the samples it did classify, it achieved 92\% precision and 95\% selective accuracy. These results show that the ranked binary precision approach remains robust even when out-of-scope samples dominate the test set, and can provide practitioners with high-confidence attributions in a large percentage of cases. 

\subsection{Operational Requirements for APT Malware Attribution}

Attributing a malware sample to an APT group requires that the sample contain sufficient distinguishing features to support that attribution. This applies both to human analysts and to automated systems. Consequently, some samples will inevitably lack enough evidence to be confidently attributed to a specific APT group.
Similarly, there will be groups for which an effective classifier cannot be built. This may occur because too few samples are available, because samples from the group do not share sufficiently consistent features, or because the group is new or has not yet been identified. In practice, attribution systems must therefore operate on a mixture of in-scope samples from groups represented during training and out-of-scope samples from groups the classifier has not been trained to recognise. Since most published models are trained on a limited number of APT groups (typically 16 or fewer APT groups), they are likely to produce incorrect classifications when deployed in operational environments.

For automated attribution to be useful to defenders, the system must communicate uncertainty rather than forcing a classification for every sample. When a sample cannot be classified reliably, abstention is preferable to an incorrect attribution, since misleading classifications can erode analyst trust and lead to inappropriate defensive actions. 
This distinguishes our approach from prior APT malware attribution systems, which do not evaluate out-of-scope behaviour, do not demonstrate comparable rejection performance under similarly challenging conditions, or have poor precision. By combining high-precision attribution with explicit abstention, the ranked binary approach provides a stronger basis for operational deployment than prior approaches. 

\subsection{Limitations}
A key limitation of this work is the limited availability of large, high-quality, group-labelled APT malware datasets. This constrained both model tuning and the extent to which the models could be evaluated across diverse APT groups. Our experiments only used the ADAPT feature extraction pipeline~\cite{SecPriv/adapt:Samples}. 
However, the ranked binary methodology is independent of the specific feature extraction technique: different features may alter the strength of the underlying classifiers and the multi-class baseline, but threshold-based abstention should still allow the operating point to be shifted towards higher precision.

There are also few directly comparable studies that evaluate out-of-scope behaviour in APT malware attribution. This limits the extent to which our results can be compared against prior work, particularly under large-scale out-of-scope conditions. 
As a result, while our experiments provide evidence of improved resilience to out-of-scope samples, further evaluation on additional malware datasets and feature extraction pipelines is needed to fully assess generalisability.

\begin{table*}[!h]
\caption{Comparison against published malware attribution approaches}
  \centering
  \label{tab:approach comparison}
  \begin{tabular}{llll}
    \hline
    Approach & Classifier Type & Tested Against Out-of-scope Groups & Known/Expected Behaviour \\
    \hline
    Lee et al.  \cite{Lee2022MalwareAnalysis} & Multi-class & Not tested & Misclassified\\
    Wang et al. \cite{Wang2022APTShapelets} & Multi-class & Not tested & Misclassified\\
    Kida et al. \cite{Kida2023Nation-StateHashing} & Multi-class & Not tested &Misclassified\\
    Saha et al. \cite{Saha2024ADAPTFiles} & Multi-class & Not tested & Misclassified\\
    Rosenberg et al. \cite{Rosenberg2017DeepAPT:Networks}, \cite{Rosenberg2018End-to-EndMalware} & Multi-class & Not tested & Misclassified\\
    Haddadpajouh et al. \cite{Haddadpajouh2020MVFCC:Attribution} & Consensus Clustering & Not tested & Misclassified\\
    Li et al. \cite{Li2021AttributionTechniques} & Multi-class & Not tested & Misclassified\\
    Xu et al. \cite{Xu2021AnLightGBM} & Multi-class & Not tested & Misclassified\\
    Wang et al. \cite{Wang2021ExplainableTechniques} & Multi-class & Not tested & Misclassified\\  
    Han et al. \cite{Han2021APTMalInsight:Framework} & Multi-class & Not tested & Unclear \\
    Hong et al. \cite{Hong2019MalwareGroups} & Multi-class &  Not tested & Misclassified\\
    Liu et al. \cite{Liu2019Functions-basedAnalysis}  & Multi-class & Not tested & Misclassified\\
    Rani et al. \cite{Rani2024GenesisAttribution} & Multi-class Soft voting  & Tested & Detection\\
    Sun et al. \cite{Sun2025MGAP3:MalwarePre-Training} & Multi-class & Not tested & Unclear \\
    Shenderovitz et al. \cite{Shenderovitz2024Bon-APT:Calls} & Multi-class & Not tested & Misclassified \\
    Tang et al. \cite{Tang2023APTFramework} & Multi-class & Not tested & Misclassified \\
    Tang et al. \cite{Tang2023DeepAnalysis} & Multi-class & Not tested & Misclassified\\
    \hline
    Our approach & Multiple Binary & Tested & Detection \\
    \hline
   \end{tabular}
\end{table*}

\section{Related work}\label{sec:related work}
Prior work on APT malware attribution differs primarily in two respects: the features extracted from malware samples and the modelling approach used to perform attribution. Existing studies have considered both static and dynamic sources of evidence. Static approaches analyse properties of the malware without execution. For example, Wang et al.~\cite{Wang2022APTShapelets} extracted shapelets, i.e., short discriminative subsequences from time-series representations. Lee et al.~\cite{Lee2022MalwareAnalysis} used the frequency of loaded modules, while Kida et al.~\cite{Kida2023Nation-StateHashing} combined similarity hashing of raw samples with vector embeddings. Saha et al.~\cite{Saha2024ADAPTFiles} combined several static analysis tools with semantic embeddings. Other static approaches include control flow graph features~\cite{Liu2019Functions-basedAnalysis}, tactics, techniques, and procedures~\cite{Rani2024GenesisAttribution}, and features derived from disassembled code~\cite{Sun2025MGAP3:MalwarePre-Training,Tang2023APTFramework,Tang2023DeepAnalysis}.

Dynamic approaches instead extract features from malware behaviour observed during execution. Several studies use reports generated by Cuckoo Sandbox. Rosenberg et al.~\cite{Rosenberg2017DeepAPT:Networks,Rosenberg2018End-to-EndMalware} used raw sandbox reports as features, while the work in \cite{Haddadpajouh2020MVFCC:Attribution} divided reports into multiple views and applied fuzzing. Li et al.~\cite{Li2021AttributionTechniques} and Xu et al.~\cite{Xu2021AnLightGBM} applied statistical methods to the terms appearing in the reports. Wang et al.~\cite{Wang2021ExplainableTechniques} proposed a hybrid approach that combines Cuckoo Sandbox reports with static analysis, while \cite{Hong2019MalwareGroups} used behavioural artefacts such as APIs, network activity, registry keys, files, and mutexes.

Feature engineering is not the focus of this work. We therefore adopt the static feature extraction method proposed by Saha et al.~\cite{Saha2024ADAPTFiles}. Since they do not report attribution metrics for their group-labelled dataset, we cannot directly compare against their attribution results. Instead, our focus is on the attribution strategy applied after feature extraction, particularly whether the classifier can avoid assigning samples to a known group when the evidence is insufficient.

Most prior work focuses on feature extraction and then applies established supervised machine learning algorithms for attribution. Some exceptions use alternative modelling paradigms. Saha et al.~\cite{Saha2024ADAPTFiles} used agglomerative clustering, and Haddadpajouh et al.~\cite{Haddadpajouh2020MVFCC:Attribution} proposed a fuzzy consensus clustering model. The work in \cite{Sun2025MGAP3:MalwarePre-Training} used a Perceiver IO encoder, \cite{Tang2023APTFramework,Tang2023DeepAnalysis} used CNN-based models with attention, and \cite{Hong2019MalwareGroups} represented malware attribution using a bipartite graph.

Our work differs from these approaches in its treatment of attribution as a selective classification problem. Rather than relying on a single closed-set multi-class classifier, we train a set of binary classifiers, with each classifier selected and tuned for a specific APT group. These classifiers are then ranked and combined to perform attribution while allowing samples to remain unclassified when no classifier provides sufficient evidence. This design enables explicit abstention, which is essential when samples may originate from out-of-scope APT groups.

This distinction has practical implications. In operational settings, APT attribution systems are unlikely to encounter only samples from groups represented during training. They may also receive samples from new, underreported, or insufficiently sampled groups. Without a mechanism for abstention, these samples are likely to be forced into one of the known classes, producing unsupported attributions that may mislead defenders rather than assist them.

Laurenza et al. \cite{Laurenza2018MalwareActivities} proposed a malware triage method that separates samples likely to be APT-related from those that are not. Their work shares several design goals with ours, including an emphasis on precision and the use of one binary classifier per APT group. However, their objective is triage rather than attribution: they do not combine the binary classifiers into an APT attribution classifier, nor do they evaluate attribution performance. In addition, they train all binary classifiers using the same techniques, whereas our approach selects and tunes classifiers independently for each APT group.

Rani et al.~\cite{Rani2024GenesisAttribution} treat APT malware attribution as an open-set problem and therefore provide the closest malware-based comparison for out-of-scope behaviour. Their approach identifies 94\% of out-of-scope samples when 43\% of samples are out of scope. In comparison, our ranked binary precision configuration rejects 92\% and 94\% of out-of-scope samples when 42\% and 87\% of samples are out of scope, respectively. However, Rani et al. report 81\% macro precision during in-scope testing, compared with 96--98\% for our in-scope ranked binary configurations, and do not report selective precision during out-of-scope testing. This limits direct comparison of classification quality under abstention.
They also calculate their threshold for highest precision against the known test set, rather than the known training set, which may bias the results.
Perry et al.~\cite{Perry2019NO-DOUBT:Reports} also consider rejection of previously unseen actors, but their work is based on threat intelligence reports rather than malware samples, so it is relevant to out-of-scope rejection but not directly comparable to APT malware attribution.

Table~\ref{tab:approach comparison} summarises how prior approaches differ in classifier type and whether they evaluate performance in the presence of out-of-scope groups. For approaches that do not report such an evaluation, we describe the expected closed-set behaviour based on whether the method includes an explicit abstention or rejection mechanism. 
In addition to this methodological distinction, our experimental results compare the ranked binary classifiers against previously published models evaluated on the same dataset. As shown earlier in Table~\ref{tab:combined-binary-classifiers}, the ranked binary approach improves precision over prior work while retaining competitive overall performance.

\section{Conclusion and future work}\label{sec:conclusion and future work}
Advanced persistent threats continue to pose a significant challenge to defenders, and early attribution can help guide more effective defensive action~\cite{Advanced2025}. However, practical APT malware attribution is difficult because operational environments are likely to contain samples from groups that were not represented during training. Existing solutions are therefore forced to assign such samples to known groups, increasing the risk of misleading attributions.

In this paper, we presented a ranked binary classification approach for high-precision APT malware attribution. Our method trains and tunes one binary classifier per APT group, ranks the classifiers according to validation performance, and applies them sequentially. This allows the classifier to abstain when no classifier provides sufficient evidence for attribution.

Our experiments show that the approach improves precision over previously published results on the same APT Malware dataset~\cite{Cyber-research/APTMalware:Samples,Xu2021AnLightGBM,Kida2023Nation-StateHashing}. In the most challenging evaluation, where 87\% of test samples came from 60 out-of-scope APT groups, the ranked binary precision classifier rejected 94\% of out-of-scope samples while achieving 92\% precision and 95\% selective accuracy on the samples it classified.
These results show that explicit abstention is critical for practical APT attribution. By prioritising high-confidence classifications and withholding attribution when evidence is insufficient, the ranked binary approach reduces unsupported attributions and provides a more operationally suitable alternative to the current state of the art. 

In future work, we will evaluate whether the ranked binary methodology provides similar benefits with other static, dynamic, or hybrid feature representations beyond the ADAPT feature extraction pipeline and on larger, evolving APT malware datasets. We will also explore alternative operating points, including post-training per-group classifier selection and per-group thresholding within multi-class classifiers, while accounting for the difficulty of calibrating scores across classes and classifier types. Finally, we will study the operational severity of different misclassifications, since errors between related APT groups may have different defensive consequences from errors involving unrelated groups.

\section*{Acknowledgment}
This work was supported by the EPSRC and MOD Centre for Doctoral Training in Complex Integrated Systems for Defence and Security [EP/Y034848/1].

\bibliographystyle{IEEEtran}
\bibliography{references}

\section{Further Details on Model Building}
\label{sec:model building}

The stratified split was important because some groups contain few samples, and a non-stratified split could leave too few samples from a group in either set. To avoid reducing the training data further, we used stratified cross-validation rather than a separate validation set. Multi-class models used five folds. For binary classifiers, the validation strategy was selected dynamically based on the proportion of positive samples. Where possible, we used stratified $k$-fold cross-validation with three to five folds, selecting the largest number of folds that retained at least 10 positive samples per fold. Otherwise, we used a stratified shuffle split.

Hyper-parameter search grids were also selected dynamically. Multi-class models used a fixed grid for each model type, while binary classifiers used grids based on the proportion of positive samples: below 0.02\%, between 0.02\% and 0.1\%, and above 0.1\%. We sampled 40 combinations per classifier, which provided a good balance between search coverage and computational cost.
After hyper-parameter tuning, we applied stratified cross-thresholding to select decision thresholds for the binary classifiers. This used the same fold-selection criteria as cross-validation, but with a different random seed. We selected the lowest threshold that achieved the best target score, avoiding unnecessarily high thresholds that reduced recall when a lower threshold achieved the same precision.

For each APT group, we selected both a \emph{high-F1 classifier} and a \emph{high-precision classifier}. High F1 classifiers were selected by F1 score, with precision as a tiebreaker. High precision classifiers were selected by precision, with recall as a tiebreaker. If a perfect classifier was found, the search was terminated early. The held-out test set was then used for the first time to generate the final metrics.

All model tuning was carried out on a Dell Precision 7680 laptop, with 32GB of RAM and a 13th Gen Intel(R) Core(TM) i7-13850HX (2.10 GHz) processor running Windows 11 Enterprise.
When tuning both the F1 and precision binary classifiers over the APT Malware dataset the complete set was built in approximately 8.7 hours. However, if this was limited to just Extra Trees and Hist Gradient Boosting models, which were selected in all but one case, this reduced to 4.6 hours.

\section{Sample Sets}
\label{sec:Sample Sets}
This section contains high level statistics of the different datasets used in these experiments.
We provide in Table \ref{tab:APTMalware dataset} the APT Malware dataset's 12 groups and their number of samples.

\begin{table}[h!]
  \caption{Summary of the APT Malware dataset \cite{Cyber-research/APTMalware:Samples}}
  \label{tab:APTMalware dataset}
  \centering
  \begin{tabular}{l|rr}
    \hline
    APT Group & Freq. & \% \\
    \hline
    APT 1 & 405 & 11\%\\
    APT 10 & 244 & 7\%\\
    APT 19 & \textbf{32} & \textbf{1}\%\\
    APT 21 & 106 & 3\%\\
    APT 28 & 214 & 6\%\\
    APT 29 & 281 & 8\%\\
    APT 30 & 164 & 5\%\\
    Dark Hotel & 273 & 8\%\\
    Energetic Bear & 132 & 4\%\\
    Equation Group & 395 & 11\%\\
    Gorgon Group & \textbf{961} & \textbf{28}\%\\
    Winnti & 387 & 11\%\\
    \hline
   \end{tabular}
\end{table}
We introduce in Table \ref{tab:Top12CombinedGroups}, the combined dataset's top 12 groups and their number of samples. We then provide in Tables \ref{tab:Top20CombinedGroups} and \ref{tab:Top25CombinedGroups} correspondingly the combined dataset's top 20 groups and 25 groups  and their number of samples.
\begin{table}[!ht]
  \caption{Top 12 Combined Groups}\label{tab:Top12CombinedGroups}
  \centering
  \begin{tabular}{l|rr|rr}
 \hline
    & \multicolumn{2}{c|}{Training} & \multicolumn{2}{c}{Test} \\
    APT Group & Freq. & \% & Freq. & \% \\
 \hline
    APT 1 & 348 & 10\% & 87 & 1\%\\
    APT 10 & 195 & 5\% & 49 & 1\%\\
    APT 28 & 210 & 6\% & 53 & 1\%\\
    APT 29 & 292 & 8\% & 74 & 1\%\\
    APT 36 & 220 & 6\% & 55 & 1\%\\
    APT 41 & 440 & 12\% & 110 & 2\% \\
    Dark Hotel & 252 & 7\% & 64 & 1\%\\
    Equation Group & 316 & 9\% & 79 & 1\%\\
    Gorgon Group & 768 & 21\% & 192 & 3\%\\
    Lazarus Group & 221 & 6\% & 56 & 1\%\\
    Turla & 138 & 4\% & 35 & 1\%\\
    WizardSpider & 200 & 6\% & 50 & 1\%\\
    60 Out-of-scope Groups & & & \textbf{5615} & \textbf{87}\%\\
\hline
   \end{tabular}
\end{table}

\begin{table}[!hb]
  \caption{Top 20 Combined Groups}\label{tab:Top20CombinedGroups}
  \centering
  \begin{tabular}{l|rr|rr}
    \hline
    & \multicolumn{2}{c|}{Training} & \multicolumn{2}{c}{Test} \\
    APT Group & Freq. & \% & Freq. & \% \\
    \hline
    APT 1 & 348 & 8\% & 87 & 5\%\\
    APT 10 & 195 & 5\% & 49 & 3\%\\
    APT 2 & 72 & 2\% & 19 & 1\% \\
    APT 21 & 81 & 2\% & 21 & 1\% \\
    APT 28 & 210 & 5\% & 53 & 3\%\\
    APT 29 & 292 & 7\% & 74 & 4\%\\
    APT 30 & 138 & 3\% & 35 & 2\%\\
    APT 36 & 220 & 5\% & 55 & 3\%\\
    APT 41 & 440 & 10\% & 110 & 6\% \\
    Dark Hotel & 252 & 6\% & 64 & 3\%\\
    Dragonfly & 112 & 3\% & 28 & 2\%\\
    Equation Group & 316 & 7\% & 79 & 4\%\\
    Fin 7 & 68 & 2\% & 17 & 1\%\\
    Gameredon & 116 & 3\% & 29 & 2\%\\
    Gorgon Group & 768 & 18\% & 192 & 10\%\\
    Lazarus Group & 221 & 5\% & 56 & 3\%\\
    TA505 & 100 & 2\% & 25 & 1\% \\
    Turla & 138 & 3\% & 35 & 2\%\\
    WizardSpider & 200 & 5\% & 50 & 3\%\\
    53 Out-of-scope Groups & & & \textbf{774} & \textbf{42}\%\\
  \hline
   \end{tabular}
\end{table}

\begin{table}[!ht]
  \caption{Top 25 Combined Groups}\label{tab:Top25CombinedGroups}
  \centering
  \scalebox{1}{
  \begin{tabular}{l|rr|rr}
    \hline
    & \multicolumn{2}{c|}{Training} & \multicolumn{2}{c}{Test} \\
    APT Group & Freq. & \% & Freq. & \% \\
    \hline
    APT 1 & 348 & 8\% & 87 & 6\%\\
    APT 10 & 195 & 4\% & 49 & 3\%\\
    APT 12 & 47 & 1\% & 12 & 1\% \\
    APT 2 & 72 & 2\% & 19 & 1\% \\
    APT 21 & 81 & 2\% & 21 & 1\% \\
    APT 27 & 35 & 1\% & 21 & 1\% \\
    APT 28 & 210 & 5\% & 53 & 3\%\\
    APT 29 & 292 & 6\% & 74 & 5\%\\
    APT 30 & 138 & 3\% & 35 & 2\%\\
    APT 36 & 220 & 5\% & 55 & 3\%\\
    APT 41 & 440 & 10\% & 110 & 7\% \\
    Blind Eagle & 48 & 1\% & 13 & 1\% \\
    Dark Hotel & 252 & 6\% & 64 & 4\%\\
    Deep Panda & 36 & 1\% & 9 & 1\%\\
    Dragonfly & 112 & 2\% & 28 & 2\%\\
    Equation Group & 316 & 7\% & 79 & 5\%\\
    Fin 7 & 68 & 1\% & 17 & 1\%\\
    Gameredon & 116 & 3\% & 29 & 2\%\\
    Gorgon Group & 768 & 17\% & 192 & 12\%\\
    Kimusky & 53 & 1\% & 14 & 1\%\\
    Lazarus Group & 221 & 5\% & 56 & 4\%\\
    Naikon & 53 & 1\% & 14 & 1\%\\
    TA505 & 100 & 2\% & 25 & 2\% \\
    Turla & 138 & 3\% & 35 & 2\%\\
    WizardSpider & 200 & 4\% & 50 & 3\%\\
    47 Out-of-scope Groups & & & \textbf{419} & \textbf{27}\%\\
   \hline
   \end{tabular}
   }
\end{table}

\section{Conceptual Metrics Descriptions}\label{sec:Conceptual Metrics Descriptions}
We provide a conceptual overview of the metrics used in Table \ref{tab:standard metrics}.

\begin{table}[!t]
\caption{Standard Metrics: A higher value is better for all metrics used}
\label{tab:standard metrics}
\centering
\scalebox{1}{
\begin{tabular}{lll} 
\hline
Metric & Short Form & Brief Conceptual Description \\
\hline
Precision & Prec & How often a binary classification is correct\\
Recall & Rec & How often a binary class is correctly identified\\
F1 & F1 & The balance between precision and recall\\
\hline
Precision (weighted) & Prec & How often a classification is correct, averaged over all samples\\
Recall (weighted) & Rec & How often a class is correctly identified, averaged over all samples\\
F1 (weighted) & F1 & The balance between weighted precision and recall\\
\hline
Precision (macro) & Prec(m) & How often a classification is correct, averaged over all classes\\
Recall (macro) & Rec(m) & How often a class is correctly identified, averaged over all classes\\
F1 (macro) & F1 (m) & The balance between macro precision and recall\\
\hline
Selective Precision (weighted) & S-Prec & Weighted precision for samples where the model makes a classification\\
Coverage & Cov & The percentage of samples a classification was made for \\
\hline
In-scope Coverage & Cov(is) & Percentage of in-scope samples that a classification was made for\\
Weighted Effective Recall (In-scope) & Rec(is) & Weighted recall calculated across in-scope samples\\
Macro Effective Recall (In-scope) & mRec(is) & Macro Recall calculated across in-scope samples\\
\hline
Selective Accuracy & S-Acc & When a classification was made, what percentage of classifications were correct \\
Selective weighted F1 & S-F1 & The balance between selective weighted precision and recall\\
\hline
Rejection Rate for OOS & OOS-RR & The percentage of out-of-scope samples for which a classification is not made\\
\hline
\end{tabular}
}
\end{table}

\end{document}